\documentstyle{mn}

\input psfig

\font\twelvei = cmmi10 scaled\magstep1
       \font\teni = cmmi10 
\font\mbf = cmmib10 scaled\magstep1
       \font\mbfs = cmmib10 \font\mbfss = cmmib10 scaled 833
\font\msybf = cmbsy10 scaled\magstep1
       \font\msybfs = cmbsy10 \font\msybfss = cmbsy10 scaled 833

\font\msybf = cmbsy10 scaled\magstep1
       \font\msybfs = cmbsy10 \font\msybfss = cmbsy10 scaled 833
       \def\bmit{\fam9 }
       \def\bmsy{\fam10 }

\def\etal{{\it et al.\ }}
\def\eg{{\it e.g.\ }}
\def\ie{{\it i.e.\ }}
\def\lsim{\mathrel{  
	\raise0.3ex\hbox{$<$}\kern-0.75em{\lower0.65ex\hbox{$\sim$}}}}
\def\gsim{\mathrel{
	\raise0.3ex\hbox{$>$}\kern-0.75em{\lower0.65ex\hbox{$\sim$}}}}
\def\kms{\mbox{\rm\,km\,s$^{-1}$}}      

\def\sun{\odot}

\title[Cluster Accretion Flows]
{Accreting Matter around Clusters of Galaxies:\\
   One-Dimensional Considerations}

\author[D.~Ryu and H.~Kang]
{Dongsu Ryu$^{1}$ and Hyesung Kang$^{2}$\\
$^1$Dept.~of Astronomy \& Space Sci., Chungnam National University, 
    Daejeon 305-764, Korea;
    ryu@sirius.chungnam.ac.kr\\
$^2$Dept.~of Earth Sciences, Pusan National University,
    Pusan 609-735, Korea;
    kang@astrophys.es.pusan.ac.kr}

\date{Accepted 1996 ** **; Received 1996 ** **}

\begin{document}

\maketitle

\begin{abstract}

During the formation of the large scale structure of the Universe,
matter accretes onto high density peaks. 
Accreting collisionless dark matter (DM) forms caustics around them,
while accreting collisional baryonic matter (BM) forms accretion shocks.
The properties of the accreting matter depend upon the power spectrum
of the initial perturbations on a given scale as well as the
background expansion in a given cosmological model.
In this paper, we have calculated the accretion of DM particles in
one-dimensional spherical geometry under various cosmological models
including the Einstein-de Sitter universe, the open universe with
$\Omega_o<1$, and the flat universe with $\Omega_{\Lambda}=1-\Omega_o$.
A density parameter in the range $0.1\le \Omega_o \le 1$ has been
considered.
The initial perturbation characterized by a point mass at the origin
has been considered.
Since the accretion shock of BM is expected to form close to the first
caustic of DM, the properties of the accreting BM are common with those of
the DM.
Hence, the accretion calculations with DM particles have been used to
find the position and velocity of the accretion shock and the cluster
mass inside it.
The average temperature of BM has been estimated by adopting simplifying
assumptions.
The velocity of the accreting BM around clusters of a given temperature
is smaller in a universe with smaller $\Omega_o$, but only by up to
$\sim24\%$ in the models with $0.1\le \Omega_o \le 1$.
Thus, it would be difficult to use that quantity to discriminate among the
cosmological models.
However, the accretion velocity around clusters of a given mass or 
a given radius depends more sensitively on the cosmological models.
It is smaller in a universe with smaller $\Omega_o$ by up to
$\sim41\%$ and $\sim65\%$, respectively.
So, it can provide a better signature of the background expansion for  
different cosmological models. 
Although the existence of the caustics and the accretion shocks may not
be confirmed by direct x-ray observations, the infalling warm gas of
$10^4-10^5$K upstream of the shocks may be observed as the absorption
systems of quasar emission lines. 
According to this study, the suggestion made by Kang, Ryu, \& Jones
(1996) that the large scale accretion shocks around clusters of galaxies
can serve as possible acceleration sites of ultra high energy cosmic rays
above $10^{18}$ eV remains plausible in all viable cosmological models.

\end{abstract}


\section{INTRODUCTION}

Recent satellite x-ray observations have made clusters of galaxies
increasingly important to cosmology as a probe into the large
scale structure of the universe.
Being massive and rare, their abundance in the local and distant
universe carries vital information on the initial density fluctuations
and the matter content of the universe (Lubin \etal 1995;
Eke, Cole, \& Frenk 1996).
Also being relatively young dynamically, the details of their structures
can provide us with some signatures left over from the formation epoch
as well as information on the background cosmology
(Crone, Evrard, \& Richstone 1994; Navarro, Frenk, \& White 1995;
Tsai \& Buote 1995).
A general consensus seems to be that the standard cold dark matter 
model (\ie, $\Omega=1$ and $h=0.5$) normalized to the COBE DMR measurement
of the anisotropies in the cosmic background radiation (\ie,
$\sigma_8>1$) has serious difficulties in explaining the observed
properties of x-ray clusters, such as the cluster abundance (Kang \etal 1994),
the baryon fraction in clusters (Lubin \etal 1995), and the contribution
of cluster emission to the x-ray background (Kang \etal 1994;
Kitayama \& Suto 1996).
On the other hand, an open CDM model or a flat CDM model with a cosmological
constant with a smaller value of $\sigma_8$ seems more
consistent with many observations, and so has become popular recently
(Cen \& Ostriker 1994; Ostriker \& Steinhardt 1995).

Although the application of the Press-Schechter formalism
(Press \& Schechter 1974) to the hierarchical clustering model of cluster
formation has been successful in getting good agreement with the results
of $N$-body simulations (Lacey \& Cole 1994),  this semi-analytic
approach cannot include the effects of non-equilibrium hydrodynamics
that could be important during some phases in cluster evolution
(Kang \etal 1994).
Some studies suggest that the amplitude of the
density power spectrum, $\sigma_8$, can be constrained by the local
cluster abundance (Eke, Cole, \& Frenk 1996; Viana \& Liddle 1995) 
and the range of the allowed value of $\Omega_o$ can be narrowed down
by examining the evolution of cluster abundance at low redshifts
(\ie, $z<1$) (Bahcall \& Cen 1992; White, Efstathiou, \& Frenk 1993).
Despite this, the observational and theoretical/numerical
errors involved in such 
procedures seem to be too big to make any consistent predictions on 
those key cosmological parameters
(Castander \etal 1995; Luppino \& Gioia 1995; Tsai \& Buote 1995).
It is also unlikely that all crucial physics can be included
in any variants of the treatments.

The properties of x-ray clusters, other than the abundance and the baryon
fraction mentioned above, which have been investigated for various
cosmological models, include the cluster-cluster correlation function
(Bahcall \& Cen 1992), the density and velocity profiles
(Crone, Evrard, \& Richstone 1994), and the degree of substructures
in the intracluster medium (ICM) (Tsai \& Buote 1995).
The latter two, associated with the internal structure of clusters,
are closely related with recent mergers and accretion onto the cluster
mass scale ($\sim 8 h^{-1}$Mpc), so dependences on $\Omega$, $n$
(the power-law index of the initial power spectrum), and $\sigma_8$ could
be degenerate.
In addition to that, statistical treatments of observed data of those
quantities have a harder time discriminating clearly among 
different cosmological models compared to the statistics of the cluster
abundance.
Even though there are tantalizing possibilities that the various properties
of x-ray clusters can indeed be used to unveil the fundamental nature 
of the universe, further improvements by a factor of at least a few 
in both theoretical and observational fronts seem to be required in
order to make some solid predictions. 

In this paper, we have examined one more physical property
associated with x-ray clusters, the accretion flow infalling
toward the clusters. 
Matter accretes onto the high density peaks continuously 
throughout the history of the universe.
Its rate on the scale of the cluster mass will be determined by the
initial density fluctuations and the background cosmology. 
Fluctuations continue to grow in the $\Omega=1$ universe, while they
stop growing at a redshift $z\sim\Omega_o^{-1}-1$ in a universe
with small $\Omega_o$ (Peebles 1993).
On the other hand, the accretion on larger scales gets stronger for 
a smaller $n$ for a scale-free power spectrum.
Crone, Evrard, \& Richstone (1994) examined the radial velocity profile of
the accretion flow through $N$-body simulations of different 
cosmological models with scale-free power spectra.
They found that the density profile is flatter and the accretion regime 
is stronger for higher $\Omega_o$ and smaller $n$.

If the initial density perturbation is scale-free, the accretion of both
baryonic matter and dark matter in the $\Omega_o=1$ universe can be
described semi-analytically with a self-similar  solution
(Fillmore \& Goldreich 1984; Bertschinger 1985).
But in a universe with $\Omega_o\not=1$, the self-similar solution
is not possible, since, in addition to the time scale marking the transition
of the initial perturbation to the nonlinear regime, another time scale
such as the cosmic time $t_{1/2}$ when $\left|1-\Omega\right|=1/2$ enters 
the problem.

Here, we use a one-dimensional spherical $N$-body code, which can follow
the evolution of collisionless dark matter particles, to find the
properties of accretion flows onto objects of cluster mass
scales in different cosmological models: 
the Einstein-de Sitter universe with $\Omega_o=1$, the low-density,
open universe with $\Omega_o<1$, and the low-density, flat universe
with $\Omega_o<1$ and $\Omega_{\Lambda}=1-\Omega_o$ (from non-zero
cosmological constant, $\Lambda$).
While most previous studies on accretion focused on the density 
profile inside the collapsed objects developed from infall, our
primary interests lie in the properties of spherical accretion flows
outside the objects.
In the accretion of both dark matter (DM) and baryonic matter (BM), the
collisionless DM forms caustics around the overdensity, while the
collisional BM forms accretion shocks stopping the infalling material.
In this situation, as pointed by Bertschinger (1985), the position of the
accretion shock of the gas with $\gamma=5/3$ is very close to the position
of the first caustic of the dark matter particles.
In addition, since in the region outside the shock the accretion
solution for BM is almost identical to that for DM, the infall velocity
of BM upstream of the accretion shock can be described by that of the DM. 
Hence, hydrodynamic calculations to follow BM are not necessary if we are
interested only in the position of the accretion shocks and the properties
of the accretion flows.

Of course, one-dimensional treatments cannot include important physics
such as virialization in the central region.
Three-dimensional simulations (Kang \etal 1994;
Navarro, Frenk, \& White 1995; Evrard, Metzler, \& Navarro 1995)
showed that the gas is shock-heated to the virial temperature and
then settles into hydrostatic equilibrium (HSE) and that the shock
separates the hydrostatic central region from the infalling flow. 
However, according to Navarro, Frenk, \& White (1995), where $N$-body/SPH
simulations were used to examine the properties of x-ray clusters in
the $\Omega=1$ CDM universe, the one-dimensional, self-similar solution
of Bertschinger (1985) matches well the density and temperature profiles of
their simulated clusters except for the inner central region. 
Thus, the properties of accretion flow can be efficiently studied by 
one-dimensional calculations. 

The plan of the paper is as follows.
In \S 2 we describe the numerical simulations.
In \S 3 we present the results focusing on the velocity of the accretion
flows.
In \S 4 we discuss the implications of the results on cosmology
and on cosmic-ray acceleration by the accretion shocks.
Details such as the evolution equations and the numerical scheme are
given in the appendices.

\section{SIMULATIONS}

We follow the evolution of collisionless dark matter particles
accreting onto an initial density enhancement in an otherwise
homogeneous universe through calculations with a spherical
$N$-body code.
The equations describing the particle motion and gravity in
comoving coordinates and the expansion of the background universe can
be found, for example, in Peebles (1993) and are summarized in 
appendix A.
The details of our numerical code are described in appendix B. 

The calculations start at the time corresponding to the initial
expansion parameter $a_i=10^{-5}a_o$ with $10^4$ particles.
Initially the particles are at rest and distributed uniformly,
except for an excess mass at the origin to initiate the collapse.
So the initial perturbation can be characterized by the {\it point-mass}
perturbation
\begin{equation}
{\delta M(r)\over M(r)}={\left(M(r)\over M_o\right)^{-1}}_,
\end{equation}
where $M(r)$ is the mass inside a radius $r$ and $M_o$ is a reference 
mass.
Note that the point-mass perturbation can be approximated by the
constant power spectrum (\ie, $P(k)\approx{\rm costant}$ with $n=0$)
(Hoffman \& Shaham 1985).
Although a scale-free power spectrum with $n=-1$ would represent better
the CDM power spectrum on the cluster scale (see \eg,
Bardeen \etal 1986 for the CDM power spectrum), the constant power
spectrum would be a better approximation on scales a bit larger
than that. 
Also according to Navarro, Frenk, \& White (1995), the self-similar
solution of Bertschinger (1985) with the above initial perturbation
represents well the structures of simulated clusters in an $\Omega=1$
universe with the CDM power spectrum except for the inner central region. 
The amount of the excess mass, $M_o$, has been chosen so 
that at the present epoch about $1/3$ of the particles are placed
inside the first caustic.
It requires an excess mass corresponding to $1-2$ particles.

Note that the present value of the Hubble parameter, or
$h\equiv H_o/100{\rm~km~s^{-1}Mpc^{-1}}$, enters the problem only
through the normalization parameters such as $t_o\propto1/h$,
$L_o\propto1/h$, and ${\bar\rho}\propto h^2$ (see the appendices).
So, we can include the dependence on $h$ implicitly by
expressing the results in term of $t_oh$, $L_oh$, and
${\bar\rho}h^{-2}$ (or equivalently $M_oh$).

If we had followed the evolution of BM with $\gamma=5/3$ as well, its
accretion shock would have occurred very close to the first caustic of
DM and its accretion velocity outside the shock would have been very
similar to that of DM outside the first caustic. 
Since the accretion shock approximately separates the inner virialized 
region from the outer infalling flow, here we will consider a cluster
to be the region  inside the first caustic.
So, below, we define the radius of the first caustic, $R_{cl}$, as the
cluster radius, the mass inside $R_{cl}$, $M_{cl}$, as the total cluster
mass, and the particle velocity just outside $R_{cl}$, $v_{acc}$, as the
cluster accretion velocity.

Our numerical set-up has been designed in a way parallel to that of
Bertschinger (1985), so our numerical solution for the $\Omega=1$ case
should be identical to his self-similar solution.
Thus, the accuracy of our code has been tested against his solution.
Figure 1 shows the present phase-space distribution of particles in
the Einstein-de Sitter universe, which can be directly compared with
Figure 6 of Bertschinger (1985).
The variables used in the plot are related to those in the appendices
as follows
\begin{equation}
\lambda = {r\over r_{ta}}_,
\end{equation}
\begin{equation}
{d\lambda\over d\xi} = \left(u+{\dot a}_or\right){t_o\over r_{ta}}
-{8\over9}\lambda,
\end{equation}
where $r_{ta}$ is the present turnaround radius.
Our smoothing length in the gravitational force corresponds to
$\lambda_{sm}=0.0936$ (see appendix B).
The plot shows that the numerical solution for $\lambda\gsim\lambda_{sm}$
agrees well with the exact analytic solution.
The position of the first caustic agrees well with that of the analytic
solution and the accretion velocity outside it agrees almost exactly.
On the other hand, for $\lambda\lsim\lambda_{sm}$ the agreement in the two
solutions is not so good.
The numerical solution has given a smaller velocity than the analytic
one and the positions of caustics have not been calculated correctly,
as expected.
However, the quantities we are interested in, $R_{cl}$, $M_{cl}$, and
$v_{acc}$,
are less affected by the smoothing.
Comparison of Figure 1 with Figure 6 of Bertschinger (1985) shows that
these quantities have been calculated accurately within an error
typically less than $5\%$ or so.

\begin{figure*}
\vspace{-2.1truein}
\psfig{figure=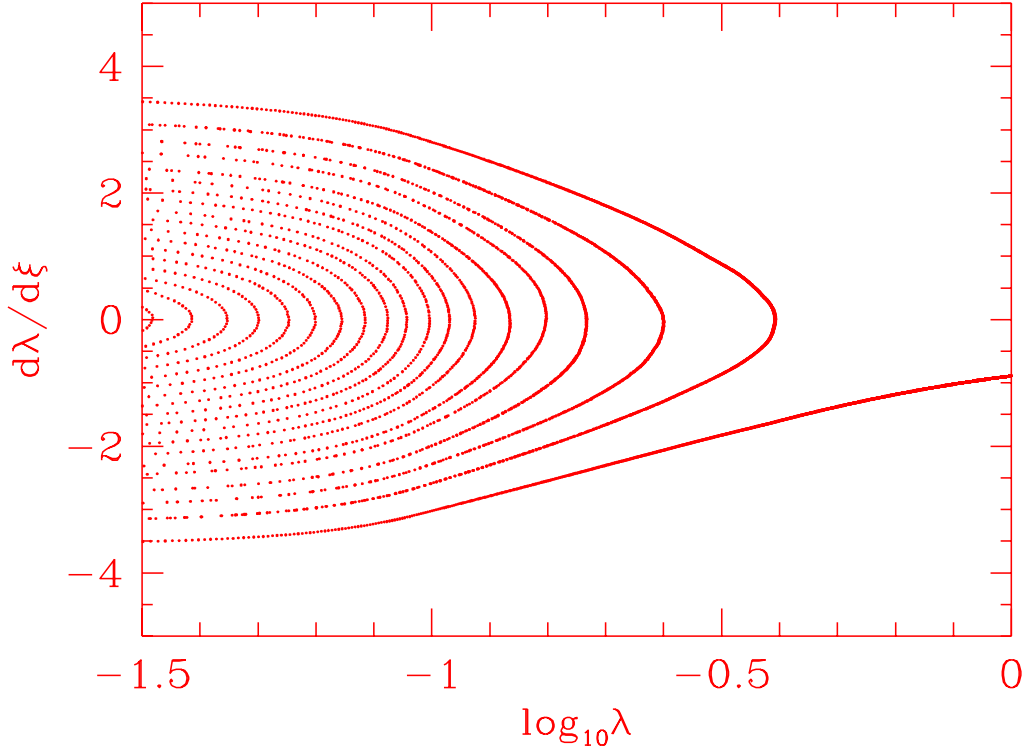,width=7.0truein,height=9.2truein}
\vspace{-2.8truein}
\caption{Present phase-space distribution of dark matter particles
from the numerical calculation in the Einstein-de Sitter universe.
It has been drawn for direct comparison with Figure 6 of Bertschinger (1985).
The relations of the variables used in the plot to those in the appendices
are $\lambda = r/r_{ta}$ and $d\lambda/d\xi = (u+{\dot a}_or)
(t_o/r_{ta})-(8/9)\lambda$.
Here, $r_{ta}$ is the present turnaround radius.}
\end{figure*}

Figure 2 shows the evolution of the mass, radius, and accretion
velocity of a cluster with the present mass of
$M_{cl}(z=0)=10^{15}h^{-1}{\rm M}_{\sun}$ as a function of redshift
in the Einstein-de Sitter universe, in the open universe with
$\Omega_o=0.2$, and in the flat universe with $\Omega_o=0.2$.
The dotted lines represent the power-law evolution of the analytic
solution in the Einstein-de Sitter universe (Bertschinger 1985);
$M_{cl} \propto (1+z)^{-1}$, $R_{cl} \propto (1+z)^{-4/3}$, and
$v_{acc}\propto (1+z)^{1/6}$.
The numerical solution in the Einstein-de Sitter universe follows
the self-similar evolution very closely.
Again, good agreement of the numerical solution with the analytic
solution indicates that our code has been able to calculate the
quantities used in this paper reliably.
But those in other model universes do not show the power-law evolution
in $(1+z)$, indicating the absence of the self-similarity.

\begin{figure}
\vspace{-0.1truein}
\psfig{figure=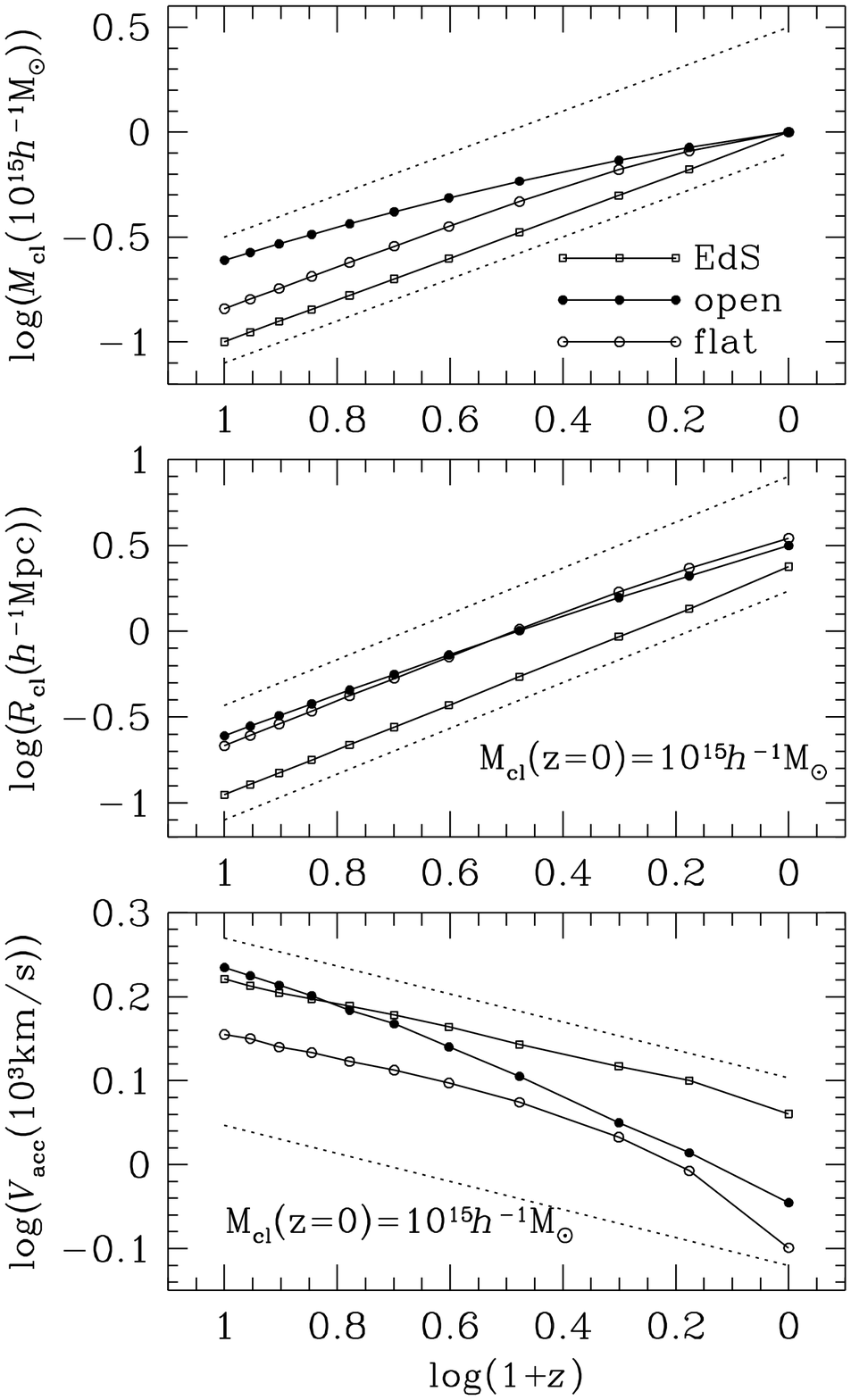,width=3.9truein,height=5.1truein}
\vspace{-0.4truein}
\caption{Evolution of the radius, mass, and accretion velocity
of a cluster with the present mass $M_{cl}(z=0)=10^{15}h^{-1}{\rm M}_{\sun}$
as a function of redshift in the Einstein-de Sitter universe (open squares),
in the open universe with $\Omega_o=0.2$ (filled circles), and in the flat 
universe with $\Omega_o=0.2$ (open circles).
The dotted lines represent the power-law evolution in the exact
self-similar solution in the Einstein-de Sitter universe (Bertschinger 1985).}
\end{figure}

\section{RESULTS}

Among the quantities we can extract from the x-ray observations of
clusters, the ICM temperature $T_{\rm x}$ (usually averaged over a core
region) is the one that can be estimated most reliably in hydrodynamic
numerical simulations and that is less prone to numerical effects
such as the problem of {\it under-resolved structure}.
That is because $T_{\rm x}$ is a conserved  quantity per unit mass
(\ie the specific thermal energy).
Although clusters have formed only recently ($z<1$) and the dynamical
time of cluster evolution is about the Hubble time, the structure of
more-or-less relaxed clusters both in observations and simulations is well
represented by the virialized region in HSE with nearly constant
temperature (Forman \& Jones 1982; Evrard, Metzler, \& Navarro 1995). 
Exceptions are the clusters that have recently undergone mergers.  
Under the assumption of an {\it isothermal sphere in HSE}, the average 
temperature is proportional to $M(<r)/r$, or to the galaxy kinetic
energy per unit mass, $\sigma_v^2$. 
So we use the ratio of the total cluster mass to radius,
$M_{cl}/R_{cl}$, as a quantity to represent the observed, average 
temperature (emission weighted). 
As shown by three-dimensional simulations (Navarro, Frenk, White 1995;
Evrard, Metzler, Navarro 1995), simulated clusters of different masses
have similar structures when they are scaled with a radius of a fixed
density contrast ($\delta = \bar \rho /\rho_{\rm crit} \gsim 200 $).
So the temperature follows the scaling law,
$T \propto M(<r_c)/r_c \propto r_c^2 $, where $r_c$ is the characteristic
radius at a fixed density contrast. 
Of course, the constant of proportionality for this scaling relation
varies for different cosmologies.

Here, we use the quantities, $M_{cl}$, $R_{cl}$, or
$M_{cl}/R_{cl}$, to denote clusters.
Note that $R_{cl}$ corresponds to the characteristic radius of
$\delta \sim 80$.
Under the set-up considered in this paper, only one of them is
independent in a given cosmological model universe, because of the
similarity in cluster structure.
According to Bertschinger (1985), the self-similar solution at the
present epoch in the Einstein-de Sitter universe gives
\begin{equation}
R_{cl} = 2.22 h^{-1} {\rm Mpc} 
{\left({ M_{cl} \over 10^{15}h^{-1} {\rm M}_{\sun} }\right)^{1/3}}_,
\end{equation}
\begin{equation}
R_{s} = 2.12 h^{-1} {\rm Mpc} 
{\left({ M_{cl} \over 10^{15}h^{-1} {\rm M}_{\sun} }\right)^{1/3}}_,
\end{equation}
\begin{equation}
T_{cl}(0.3R_s) = 6.06~{\rm keV}
{\left({ M_{cl} \over 10^{15}h^{-1} {\rm M}_{\sun} }\right)^{2/3}}_,
\end{equation}
\begin{equation}
v_{acc} = 1.31\times 10^3 {\rm km~s^{-1}} 
{\left({ M_{cl} \over 10^{15}h^{-1} {\rm M}_{\sun} }\right)^{1/3}}_.
\end{equation}
Here $R_s$ is the shock radius, and $T_{cl}$ is the temperature at
$r=0.3R_{s}$, which we may take as an average ICM temperature of the
self-similar flow.
Navarro, Frenk, \& White (1995) showed that the temperature profiles of
simulated clusters normalized with the temperature at $r_{200}$ of
$\delta=200$ can be represented within a factor of two by that of the
self-similar solution of Bertschinger (1985) in $r/r_{200} \gsim 0.3$.
In the core region within $r/r_{200}<0.3$, the temperature is
approximately isothermal.
Thus, $T_{cl}$ should be a reasonable approximation to the average 
temperature of the cluster core region.

In Figure 3 we have plotted the present $M_{cl}/R_{cl}$ for clusters
with the present mass $M_{cl}(z=0)=10^{15}h^{-1}{\rm M}_{\sun}$
(upper panel) and for clusters with the present radius
$R_{cl}(z=0)=2.5h^{-1}{\rm Mpc}$ (lower panel) in the open models
and flat models with $\Omega_o\leq1$ against the values of
$\Omega_o$.
Those with $\Omega_o=1$ correspond to the values 
in the Einstein-de Sitter universe.
Here we have chosen $R_{cl}=2.5h^{-1} {\rm Mpc}$ as a fiducial value
of the first caustic, since the typical radius of cluster core regions
is often defined as $r_{core}=0.5 h^{-1}$Mpc and the Abell radius is
$R_A=1.5 h^{-1}$Mpc. 
According to the self-similar solution of Bertschinger (1985) given above,
in the Einstein-de Sitter universe, 
\begin{equation}
{M_{cl}\over R_{cl}}=4.51\times10^{14}{{\rm M}_{\sun}\over{\rm Mpc}}
\left(M_{cl}\over10^{15}h^{-1}{\rm M}_{\sun}\right)^{2/3}
\end{equation}
for clusters with the given mass, while
\begin{equation}
{M_{cl}\over R_{cl}}=5.73\times10^{14}{{\rm M}_{\sun}\over{\rm Mpc}}
\left(R_{cl}\over2.5h^{-1}{\rm Mpc}\right)^2
\end{equation}
for clusters with the given radius.
Comparison between these values and our numerical results for
$\Omega_o=1$ case shows about $\sim 10$\% error.
Since $M_{cl}/R_{cl}$ scales as $M_{cl}/R_{cl} \propto (M_{cl}h)^{2/3}$
for clusters with different masses and $M_{cl}/R_{cl} \propto (R_{cl}h)^{2}$
for clusters with different radii, the results can be scaled accordingly. 

\begin{figure}
\vspace{-0.1truein}
\psfig{figure=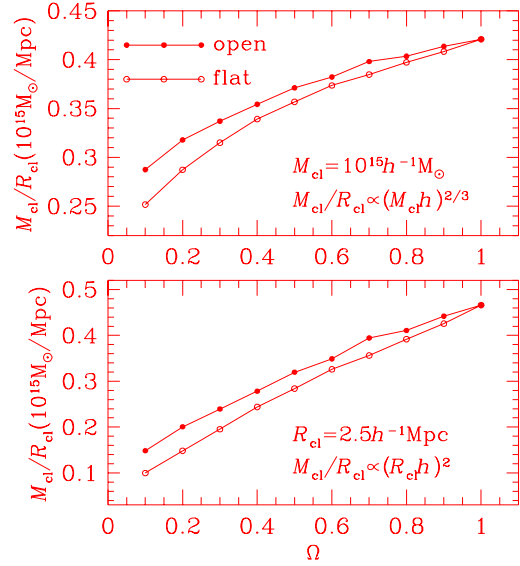,width=3.9truein,height=5.1truein}
\vspace{-1.8truein}
\caption{Present $M_{cl}/R_{cl}$ for clusters with the present mass
$M_{cl}(z=0)=10^{15}h^{-1}{\rm M}_{\sun}$ (upper panel) and for cluster
with the present radius $R_{cl}(z=0)=2.5h^{-1}{\rm Mpc}$ (lower panel)
in the open universes (filled circles) and the flat universes (open circles)
with different $\Omega_o$.
That with $\Omega_o=1$ represents the values in the Einstein-de Sitter
universe.}
\end{figure}

The figure shows that the clusters, if they have the same mass or the same
radius, are less tightly bound, and so have smaller $M_{cl}/R_{cl}$ in the
universe with smaller $\Omega_o$.
Also, the clusters are less tightly bound in the flat universe with
non-zero $\Lambda$ than in the open universe, if $\Omega_o$ is the same.
This is due to the fact the background expands faster in the flat
universe with non-zero $\Lambda$ than in the open universe.
In the open universe with $\Omega_o=0.1$, $M_{cl}/R_{cl}$ is
$\sim32\%$ and $\sim68\%$ smaller than that in the Einstein-de Sitter
universe for clusters with given mass and radius, respectively.
In the flat universe with $\Omega_o=0.1$, $M_{cl}/R_{cl}$ is
$\sim40\%$ and $\sim79\%$ smaller.
This means that clusters are less massive and less bound so the
ICM temperature is lower in low density universes than in high density
universes, if they are selected by a constant radius criterion. 

Figure 4 shows the value of $M_{cl}/R_{cl}$ as a function of redshift in
the Einstein-de Sitter universe, in the open universe with $\Omega_o=0.2$,
and in the flat universe with $\Omega_o=0.2$.
The upper panel is for clusters whose mass is 
$M_{cl}(z)=10^{15}h^{-1}{\rm M}_{\sun}$ at given $z$.
Similarly, the lower panel is for clusters whose radius is
$R_{cl}(z)=2.5h^{-1}{\rm Mpc}$ at given $z$. 
Thus, each point in the plots represent different clusters that should 
become heavier than $10^{15}h^{-1}{\rm M}_{\sun}$ or bigger in radius
than $2.5h^{-1}$ Mpc at the present epoch. 
In other words, the redshift dependences in these plots do not represent
the evolution of a particular cluster in a ``Lagrangian'' sense.
They tell us that clusters selected by either a constant mass or 
a constant radius criterion would be more tightly bound 
and so hotter in the past (at higher redshifts) than at the present epoch
in all cosmologies. 
The dotted lines represent the power-law redshift dependences 
($M_{cl}/R_{cl} \propto (1+z)$ for clusters with a constant mass, and
$M_{cl}/R_{cl} \propto (1+z)^{3}$ for clusters with a constant radius)
of the self-similar solution in the Einstein-de Sitter universe.
The numerical solution in the Einstein-de Sitter universe follows the
self-similar solutions very well.
Once again we can see that clusters are less tightly bound in lower 
density universes.
The increases of $M_{cl}/R_{cl}$ with increasing redshift in both lower
density universes are slightly less than that in the Einstein-de
Sitter universe.

\begin{figure}
\vspace{-0.1truein}
\psfig{figure=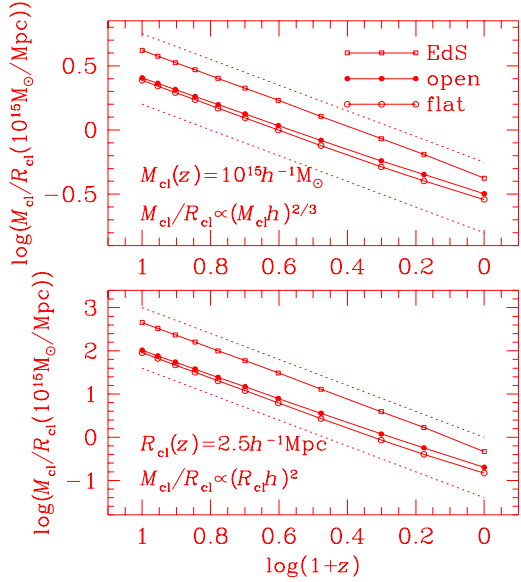,width=3.9truein,height=5.1truein}
\vspace{-1.8truein}
\caption{$M_{cl}/R_{cl}$ as a function of redshift in
the Einstein-de Sitter universe (open squares),
in the open universe with $\Omega_o=0.2$
(filled circles), and in the flat universe with $\Omega_o=0.2$
(open circles).
The upper panel is for clusters with a mass at given $z$,
$M_{cl}(z)=10^{15}h^{-1}{\rm M}_{\sun}$.
The lower panel is for clusters with a radius at given $z$,
$R_{cl}(z)=2.5h^{-1}{\rm Mpc}$.
The dotted lines represent the power-law evolution in the exact
self-similar solution in the Einstein-de Sitter universe (Bertschinger 1985).}
\end{figure}

In Figure 5 we have plotted the accretion velocity, $v_{acc}$,
for clusters with the present mass $M_{cl}(z=0)=10^{15}h^{-1}{\rm M}_{\sun}$
(the first panel), for clusters with the present radius
$R_{cl}(z=0)=2.5h^{-1}{\rm Mpc}$ (the second panel), and for clusters
with the present $M_{cl}/R_{cl}(z=0)=4\times10^{14}{\rm M}_{\sun}/{\rm Mpc}$
(the third panel) in the open and flat universes with different
$\Omega_o$'s.
Note that $M_{cl}/R_{cl}=4\times10^{14}{\rm M}_{\sun}/{\rm Mpc}$
corresponds to $T_{cl}=5.71~{\rm keV}$ in the Einstein-de Sitter
universe (see Eq.~6) (corresponding $T_{cl}$ should be slightly lower in
the low density universes).
Clusters with different masses, radii, and $M_{cl}/R_{cl}$'s have
the accretion velocity which scales as $v_{acc} \propto (M_{cl}h)^{1/3}$,
$v_{acc} \propto R_{cl}h$, and $v_{acc} \propto (M_{cl}/R_{cl})^{1/2}$.
Smaller accretion velocity in the universes with smaller $\Omega_o$
is consistent with the fact that the clusters are less tightly bound.
The accretion velocity is slight smaller in the flat universe than 
in the open universe, if $\Omega_o$ is same.
The plot shows that the accretion velocity of clusters with a given
mass or a given radius is smaller by up to $\sim41\%$ and $\sim65\%$
respectively in the flat universe with $\Omega_o=0.1$ than in the
Einstein-de Sitter universe.

\begin{figure}
\vspace{-0.1truein}
\psfig{figure=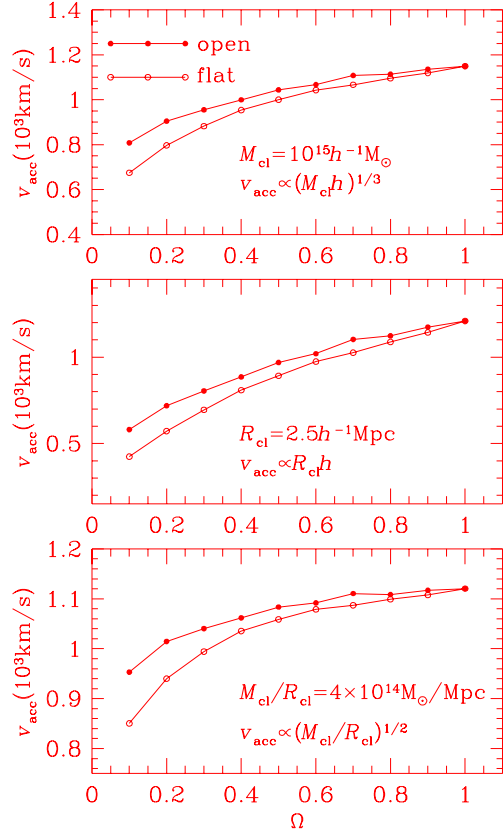,width=3.9truein,height=5.1truein}
\vspace{-0.4truein}
\caption{Present accretion velocity, $v_{acc}$, for clusters with
the present mass $M_{cl}(z=0)=10^{15}h^{-1}{\rm M}_{\sun}$
(the first panel), for clusters with the present radius
$R_{cl}(z=0)=2.5h^{-1}{\rm Mpc}$ (the second panel), and for clusters
with the present $M_{cl}/R_{cl}=4\times10^{14}{\rm M}_{\sun}/{\rm Mpc}$
(third panel) in the open universes (filled circles) and
the flat universes (open circles) with different $\Omega_o$.
That with $\Omega_o=1$ represents the values in the Einstein-de Sitter
universe.}
\end{figure}

For clusters with the same $M_{cl}/R_{cl}$, however, the difference
between different cosmological models is rather small.
In the open universe with $\Omega_o=0.1$ the accretion velocity is
$\sim15\%$ smaller than that in the Einstein-de Sitter universe,
while in the flat universe with $\Omega_o=0.1$ it is smaller
by $\sim24\%$.
This means we can expect that the clusters with the same observed
temperature can have accretion velocities smaller only up to $\sim 25 \%$
in low density universe depending on $\Omega_o$.

Figure 6 shows the accretion velocity 
for clusters with a mass of $M_{cl}(z)=10^{15}h^{-1}{\rm M}_{\sun}$
at given $z$ (first panel),
for clusters with a radius of $R_{cl}(z)=2.5h^{-1}{\rm Mpc}$
at given $z$ (second panel),
and for clusters with $M_{cl}/R_{cl}(z)=4\times10^{14}{\rm M}_{\sun}{\rm Mpc}$
at given $z$ (third panel)
in the Einstein-de Sitter universe, in the open universe with
$\Omega_o=0.2$, and in the flat universe with $\Omega_o=0.2$.
As in Fig. 4, they do not represent the evolutionary
path of a cluster, but they show the dependence of $v_{acc}$ on the
redshift in a sample of clusters selected by a constant mass, or a constant
radius, or a constant temperature criterion.
The dotted lines represent the power-law dependence on the redshift
of the self-similar solution in the Einstein-de Sitter universe,
$v_{acc} \propto (1+z)^{1/2}$ for clusters with a constant mass, 
$v_{acc} \propto (1+z)^{3/2}$ for clusters with a constant radius,
$v_{acc} \propto {\rm constant}$ for clusters with a constant $M_{cl}/R_{cl}$.

\begin{figure}
\vspace{-0.1truein}
\psfig{figure=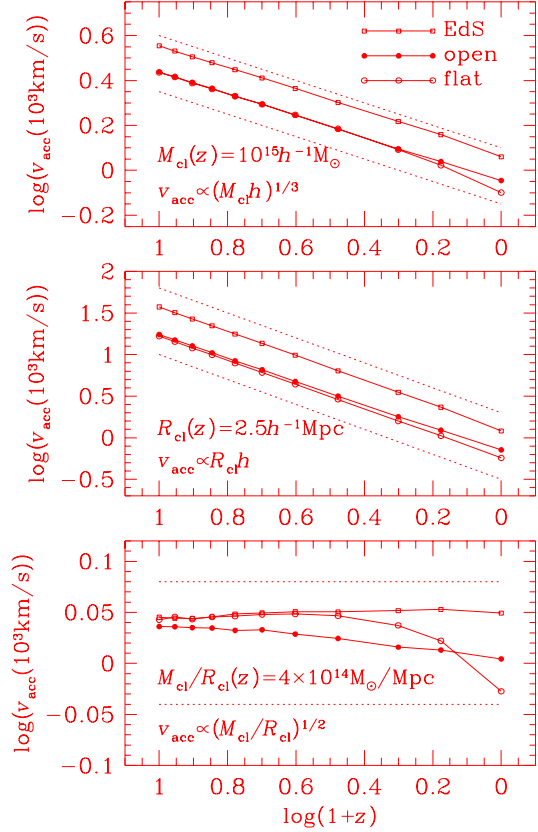,width=3.9truein,height=5.1truein}
\vspace{-0.4truein}
\caption{Accretion velocity, $v_{acc}$, as
a function of redshift in the Einstein-de Sitter universe (open squares),
in the open 
universe with $\Omega_o=0.2$ (filled circles), and in the flat universe
with $\Omega_o=0.2$ (open circles).
The first panel is for clusters with a mass at given $z$,
$M_{cl}(z)=10^{15}h^{-1}{\rm M}_{\sun}$.
The second panel is for clusters with a radius at
given $z$, $R_{cl}(z)=2.5h^{-1}{\rm Mpc}$.
The third panel is for clusters with $M_{cl}/R_{cl}$ at
given $z$, $M_{cl}/R_{cl}(z)=4\times10^{14}{\rm M}_{\sun}{\rm Mpc}$.
The dotted lines represent the power-law evolution in the exact
self-similar solution in the Einstein-de Sitter universe (Bertschinger 1985).}
\end{figure}

The first two panels show that the accretion velocity of clusters with
a constant mass or a constant radius was significantly larger in the past
than now.
On the other hand, the third panel shows that the accretion velocity
of clusters with a constant $M_{cl}/R_{cl}$ remains constant with redshift
in the Einstein-de Sitter universe due to the scale-free nature.
But it was larger at high redshifts in the model universes with
$\Omega_o<1$.
For instance, in the open universe with $\Omega_o=0.2$ the accretion
velocity was larger by $\sim 8\%$ at high redshifts, while in the flat
universe with $\Omega_o=0.2$ it was larger by $\sim 18\%$.
At high redshifts ($z>5$) the difference in $v_{acc}$ among different
cosmological models considered here reduces to only a few percents.

\section{SUMMARY AND DISCUSSION}

In this paper, we have studied the properties of spherically accreting
flows onto an initially overdense perturbation in different cosmological
model universes including the Einstein-de Sitter universe with
$\Omega_o=1$, the low-density open universe with $\Omega_o<1$,  and
the low-density, flat universe with $\Omega_o<1$ and
$\Omega_{\Lambda}=1-\Omega_o$.
According to the semi-analytic treatment of the self-similar solution
for the Einstein-de Sitter universe by Bertschinger (1985), the
accretion shock of BM with $\gamma=5/3$ forms very close to the first
caustic of DM, and the flow upstream to the accretion shock and the first
caustic is identical regardless of whether it is collisional or
collisionless.
Thus, we have assumed that the position of the accretion shock can be
approximated by that of the first caustic and the accretion velocity
of BM outside the accretion shock by that of DM outside the first caustic. 
Since self-similar solutions do not exist for $\Omega_o \ne 1$ universes,
we have used a one-dimensional, spherical, $N$-body code to study the
properties of accreting matter.

The accretion velocity onto a cluster with a given initial perturbation
has decreased as the universe expands in all cosmological models.
In the Einstein-de Sitter universe it has followed
$v_{acc} \propto (1+z)^{1/6}$ (Bertschinger 1985), while in low density
universes it decreases faster with time, especially at low redshifts
(see Figure 2).
This is related to the smaller deceleration of the expanding background.

The properties of accretion flows around the clusters of the same 
characteristic (mass, radius, or temperature) will depend on the
cosmological parameters, $\Omega_o$ and $\Lambda$, since the accretion
rate is affected by the expansion rate of the background universe. 
The accretion velocity is smaller in a lower density universe and even
smaller in a non-zero $\Lambda$ universe.
For clusters of a given mass at given redshift, the present accretion 
velocity in the low density universe with $\Omega_o\geq0.1$ is smaller
by up to $\sim41\%$ than that in the Einstein-de Sitter universe.
For clusters of a given radius at given redshift, the present accretion
velocity in the low density universe with $\Omega_o\geq0.1$ is smaller
by up to $\sim65\%$ than that in the Einstein-de Sitter universe.
Hence, if the accretion velocity of infalling matter around clusters
can be measured along with their mass and radius, then it could be 
used to discriminate the different cosmological models.

However, for clusters of a given temperature or given $M_{cl}/R_{cl}$,
the present value of the accretion velocity as well as its evolution
depend on the cosmological model rather weakly.
According to the self-similar solution of Bertschinger (1985),
in the Einstein-de Sitter universe, for clusters of a given temperature,
the accretion velocity at present is given as
$v_{acc}=1.31\times10^3{\rm km~s^{-1}}(T_{cl}/6.06 {\rm keV})^{1/2}$,
and it has been constant through the evolution of the universe.
In the universes with smaller $\Omega_o$, the present accretion
velocity is smaller by up to $\sim 24\%$ in the models considered here
with $0.1\le\Omega_o\le1$.
But this difference decreases to a few percents at high redshifts (see
Figs. 5 and 6).
Thus, it would be difficult to use the accretion velocity of clusters
with a given temperature to discriminate among the cosmological models.

However, this has an important implication for the model of the origin 
of ultra high energy cosmic rays that the acceleration of protons
in accretion shocks around clusters can contribute
significantly to the observed particle flux above $3\times 10^{18}$eV 
(Kang, Ryu, \& Jones 1996; Kang, Rachen \& Biermann 1996).
We have shown in this paper that the matter accretion and
the accretion shock around clusters are universal for all cosmological
models, although the accretion regime is less strong in lower density
universes.
The shock velocity $v_s=(4/3)v_{acc}$ for the rich clusters of $T_{cl}=10$
keV, for example, is estimated to be $2200 {\rm km~s^{-1}}$ in
the Einstein-de Sitter universe, which is enough for the model to work.
The results in this paper imply that the model might be extended to
all cosmologically viable universes of low density, since the
reduction in the shock velocity is no more than $\sim 24\%$. 

The region of clusters observed by either optical or x-ray observations
($r \lsim 1-1.5 h^{-1}$Mpc) is inside the first caustic or the accretion
shock ($R_{cl}\sim 1-3 h^{-1}$ Mpc for $T_{cl} \sim 1-10$ keV).
Although the existence of hot gas heated by the accretion shocks
has been shown clearly in most cosmological simulations including
hydrodynamics for any variants of cosmological models (Kang \etal 1994;
Cen \& Ostriker 1994; Navarro, Frenk, \& White 1995), any direct
observations of gas near the shock would not be possible at present
due to low surface brightness.
But it has been suggested that the unbound hot gas of $10^5-10^6$ K
around clusters and groups heated by the large scale accretion shocks
is a major component of the IGM which emits
mostly at soft x-ray below $1$ keV (Ostriker \& Cen 1996).
This prediction might be tested by looking for the spatial correlation 
between the soft x-ray cosmic background radiation and the observed large 
scale structure. 

On the other hand, the infalling clouds of unshocked, warm gas 
may be identified
through the absorption lines of quasars which are located inside
clusters of galaxies.
This warm low density gas of $10^4-10^5$K is photoionized by the 
diffuse radiation from the hot postshock gas and the diffuse cosmic 
background radiation. 
In some studies (\eg, Weymann \etal 1979; Foltz \etal 1986),
the C IV absorption systems of quasar emission lines with 
$ |z_{abs}-z_{QSO}| < 3000 {\rm km~s^{-1}}$ are interpreted as
clouds associated with rich clusters where the quasars reside. 
The characteristics of these systems of C IV absorbers are different from
those of the typical intervening C IV absorbers.
It was noted that the velocity difference is somewhat too large 
compared to the typical velocity dispersion of galaxies ($400-1200\kms$)
in rich clusters (Foltz \etal 1986).  
The accretion velocity, however, is a bit larger than the galaxy velocity
dispersions, since it is given by $v_{acc}= 1.31\times 10^3 {\rm km~s^{-1}}
(T_{cl}/6.06 {\rm keV})^{1/2}$ in the Einstein-de Sitter universe.
Thus it is possible that these absorption systems are in fact the 
infalling clumps of gas upstream of the accretion shock. 

In follow-up papers we will study the properties and statistics of 
three-dimensional accretion flows in simulated universes by analyzing
the data from three-dimensional hydrodynamic simulations of various
cosmological models (\eg, Kang \etal 1994; Cen \& Ostriker 1994) .
These simulations have been performed by a high-resolution, grid-based, 
Eulerian code (Ryu \etal 1993) which resolves the shock discontinuity 
in 2-3 cells and is designed specifically to handle the flows with supersonic
bulk motions.
Thus, the low density regions around caustics and shocks are well
represented in these simulations, even though the high density 
core region of clusters might be under-resolved (Kang \etal 1994).
The present one-dimensional study will provide some useful guidance for  
such studies.

\section*{acknowledgments}
The authors thank Dr. P. Biermann for directing them to the observations of
quasar absorption systems and Drs. T. W. Jones and D. H. Weinberg for
comments on the manuscript.
The work by DR was supported in part by the Basic Science Research Institute
Program, Korean Ministry of Education 1995, Project No.~BSRI-95-5408.

\appendix
\section{BASIC EQUATIONS}

In comoving coordinates, the equations to describe the evolution of
the collisionless dark matter particles are written as
\begin{equation}
{d{\bmit r}\over dt} = {1\over a}{\bmit u},
\end{equation}
and
\begin{equation}
{d{\bmit u}\over dt} = -{{\dot a}\over a}{\bmit u}-{1\over a}
{\bmsy\nabla}\phi,
\end{equation}
where ${\bmit r}$ is the comoving position, ${\bmit u}$ is the proper
peculiar velocity, $\phi$ is the proper peculiar gravitational potential,
and $a$ is the cosmic expansion parameter.
The potential is given by the Poisson equation
\begin{equation}
\nabla^2\phi = {a_o^3\over a}4\pi G\left[\rho({\bmit r})-{\bar\rho}
\right],
\end{equation}
where $\rho({\bmit r})$ is the comoving matter density,
${\bar\rho}$ is the average comoving matter density, and
$a_o$ is the present value of the the expansion parameter.

The expansion parameter and its time derivative (or the expansion
rate) are calculated as a function of time by the following equations
(see Peebles 1993).
\begin{equation}
{a\over a_o}=\left({3\over2}H_ot\right)^{2/3}, ~~~\qquad\qquad{\rm for~
Einstein-de Sitter}
\end{equation}
\begin{displaymath}
{a\over a_o}={\Omega_o\over2\left(1-\Omega_o\right)}\left(\cosh\eta
-1\right) \quad {\rm and}
\end{displaymath}
\begin{equation}
\qquad H_ot={\Omega_o\over2\left(1-\Omega_o\right)^{3/2}}\left(
\sinh\eta-\eta\right), ~~~\qquad{\rm for~open}
\end{equation}
\begin{equation}
{a\over a_o}=\left(\Omega_o\over1-\Omega_o\right)^{1/3}
\sinh^{2/3}\left({3\over2}\sqrt{1-\Omega_o}H_ot\right)_,
~~{\rm for~flat}
\end{equation}
and
\begin{equation}
\left({\dot a}\over a\right)^2=H_o^2\left[\Omega_o\left(a_o\over
a\right)^3+\Omega_R\left(a_o\over a\right)^2+\Omega_{\Lambda}\right]_.
\end{equation}
Here, $H_o$ is the present value of the Hubble parameter and $\Omega_o$,
$\Omega_R$, and $\Omega_{\Lambda}$ are constants.
The density parameter is given as
\begin{equation}
\Omega_o={8\pi G{\bar\rho}\over3H_o^2}_,
\end{equation}
which is the present average mass density in terms of the critical
density.
The parameter associated with the radius of curvature $R$ is given as
\begin{equation}
\Omega_R={1\over \left(a_oH_oR\right)^2}_,
\end{equation}
which is positive for an open universe and zero for others.
The parameter associated with the cosmological parameter $\Lambda$ is
given as
\begin{equation}
\Omega_{\Lambda}={\Lambda\over3H_o^2}_,
\end{equation}
which is positive for a flat universe and zero for others.
The three parameters give the relative contributions to the present
expansion rate by satisfying
\begin{equation}
\Omega_o+\Omega_R+\Omega_{\Lambda}=1.
\end{equation}

We have solved numerically the equations governing the evolution
of the dark matter particles [Eqs.~(A1)-(A3)], simultaneously with
the equations describing the background universe [Eqs.~(A4)-(A7)].
In the next appendix, we describe the scheme used.

\section{NUMERICAL SCHEME}

In the one-dimensional spherical code to calculate the time integration
of the equations of motion [Eqs.~(A1)-(A2)], we have adopted the second-order
accurate Lax-Wendroff scheme instead of the more popular leapfrog scheme,
since the code was originally designed to be a part of a one-dimensional
code which follows the evolution of the baryonic matter as well as that
of the dark matter in a way parallel to the three-dimensional cosmological
hydrodynamic code (Ryu \etal 1993).
So the update of the position and velocity of the particle $i$ from $n$
time step to $n+1$ is carried out by the following two steps:
\begin{equation}
r_i^{n+1/2} = r_i^n + {\Delta t^n\over2}{u_i^n\over a^n},
\end{equation}
\begin{equation}
u_i^{n+1/2} = u_i^n - {\Delta t^n\over2}{{\dot a}^n\over a^n}u_i^n
+{\Delta t^n\over 2}g_i^n,
\end{equation}
and
\begin{equation}
r_i^{n+1} = r_i^n + \Delta t^n {u_i^{n+1/2}\over a^{n+1/2}},
\end{equation}
\begin{equation}
u_i^{n+1} = u_i^n - \Delta t^n {{\dot a}^{n+1/2}\over a^{n+1/2}}
u_i^{n+1/2}+\Delta t^n g_i^{n+1/2},
\end{equation}
where $g$ denotes the gravitational force.
With the variables normalized with the present age of the universe,
$t_o$, the comoving size of the box, $L_o$, and $\bar\rho$, $g_i^n$
is calculated with
\begin{equation}
g_i^n=-{3\over8\pi}\left(H_ot_o\right)^2\Omega_o{a_o^3\over (a^{n})^2}
{\delta M(r_i^n)\over [\max(r_i^{n},\epsilon)]^2}_,
\end{equation}
where $\delta M(r_i^n)$ is the mass excess inside $r_i^n$.
Similarly, $g_i^{n+1/2}$ is calculated with the quantities at $n+1/2$.
Here $\epsilon$ is the smoothing parameter which prevents the time step
becoming too short, or prevents the particles being accelerated
anomalously for a given time step around the origin.

A particle, which is placed close to and approaches the origin, can
pass the origin in the half time step $(n+1/2)$.
Then, the absolute value of its position is used to calculate the
gravitational force at the half time step.
If a particle passes the origin after the full time step $(n+1)$,
its position and velocity are reset to their negative values.

The time step, $\Delta t$, is determined so that $\Delta a/a$ is
constant in each time step.
In all the calculations discussed in this paper, we have used $10^4$
particles and $\Delta a/a=10^{-3}$.
With these, $\epsilon=5\times10^{-2}$ in unit of $L_o$ assures that
the gravitational time scale in the core region with $r<\epsilon$ is
comfortably small compared to the time step along the whole calculation.
However, the force smoothing makes the density smoothed in the core
region with $r\lsim\epsilon$ (see \S 2).

\bsp

\end{document}